# Ultrathin layers of $\beta$-tellurene grown on highly oriented pyrolytic graphite by molecular-beam epitaxy

Jinglei Chen[1#], Yawei Dai[1#], Yaqiang Ma[2], Xianqi Dai[2†], Wingkin Ho[1], Maohai Xie[1†]

[1]*Physics Department, The University of Hong Kong, Pokfulam Road, Hong Kong*

[2]*College of Physics and Materials Science, Henan Normal University, Xinxiang, Henan 453007, China; School of Physics and Electronic Engineering, Zhengzhou Normal University, Zhengzhou, Henan 450044, China*

**Abstract**

Monolayer Tellurium (Te) or tellurene has been suggested by a recent theory as a new two-dimensional (2D) system with great electronic and optoelectronic promises. Here we present an experimental study of epitaxial Te deposited on highly oriented pyrolytic graphite (HOPG) substrate by molecular-beam epitaxy. Scanning tunneling microscopy of ultrathin layers of Te reveals rectangular surface cells with the cell size consistent with the theoretically predicted $\beta$-tellurene, whereas for thicker films, the cell size is more consistent with that of the $(10\bar{1}0)$ surface of bulk Te crystal. Scanning tunneling spectroscopy measurements show the films are semiconductors with the energy bandgaps decreasing with increasing film thickness, and the gap narrowing occurs predominantly at the valance-band maximum (VBM). The latter is understood by strong coupling of states at the VBM but a weak coupling at conduction band minimum (CBM) as revealed by density functional theory calculations.

[#]These authors contribute equally;
[†]Corresponding authors: xqdai@htu.cn (XQD, theory), and mhxie@hku.hk (MHX, experiment).

Two-dimensional (2D) materials have been the subject of intensive research in recent years. They exhibit an array of interesting properties as well as application promises in future nano-electronic, optoelectronic, spin and valley electronic devices [1-10]. Examples of widely studied 2D systems include graphene [1-3], silicene [4], phosphorene [5], borophene [6,7], stanene [8] and transition-metal dichalcogenides [9,10], covering a wide range of properties from, e.g., metals, semiconductors to insulators. Efforts are continuing to add more members to the 2D family. A recent theoretical study has suggested a group-VI element, *i.e.*, tellurium (Te), which can stabilize in the 'monolayer' form. The tellurene as is called consists of a tri-layer of Te atoms arranged in one of three configurations: 1T-$MoS_2$-like ($\alpha$-Te), tetragonal ($\beta$-Te), and 2H-$MoS_2$-like ($\gamma$-Te). The first two were found semiconducting with energy gaps of 0.75eV and 1.47 eV, respectively, which are appealing for microelectronic and optoelectronic applications [11].

At ambient conditions, crystalline tellurium (Te-I) is a semiconductor with a narrow band gap of 0.34 eV. It has a helical structure with the space group $P3_121$ or $P3_221$ depending on the chirality of Te helical chains along the *c*-axis. Atoms in each helical chain are covalently bonded whereas the chains themselves are held by van der Waals (vdW) forces and packed into a hexagonal lattice [12]. This would make it possible to fabricate Te nanostructures, for example Te nanowires and nanotubes were reportedly obtained by syntheses using different methods and techniques [13-20]. However, Te thin films particularly in the ultrathin and monolayer limit remain to be documented. At the time of preparation of this manuscript, we became aware of a recent study of epitaxial Te on graphene/6H-SiC(0001) by using molecular beam epitaxy (MBE) [21]. It was asserted that the film was a bulk-truncate of Te crystal along the [10$\bar{1}$0] direction, which could only form on graphene but not on highly oriented pyrolytic graphite (HOPG). In this Letter, we report growth of epitaxial Te on HOPG by MBE where the epifilm shows a surface lattice constant consistent with the theoretically predicted β-tellurene for ultrathin films. By a scanning tunneling microscopy and spectroscopy (STM/STS) study, we establish that the films are semiconductors and the bandgap changes with film thickness, where the change happens predominantly at the valence-band maximum (VBM), while the position of conduction-band minimum (CBM) remains almost constant. Aided by density functional theory (DFT) calculations, we attribute the latter to a strong coupling of states at the valence band edge, which has the character of the *p*-orbital of the top layer Te in the tellurene tri-layer. Interestingly, as the epifilm thickness increases, we note a change of surface cell size from $4.3 \times 5.4 \text{ Å}^2$ to $4.3 \times 6.0 \text{ Å}^2$, indicating a possible phase transformation from β-tellurene-like to bulk-like.

Te deposition on a freshly cleaved and thoroughly annealed HOPG substrate was carried out in an Omicron MBE reactor having the base pressure of ~ $2 \times 10^{-10}$ mbar. The flux of Te was generated from a standard Knudsen cell operated at 540 K. The substrate temperature was held constant at 400 K during deposition. Reflection high-energy electron diffraction (RHEED) operated at 15 keV was employed to monitor the sample surface during the deposition process and the streaky diffraction pattern signifies atomically smooth surfaces of the grown film, which is affirmed by

STM measurements at room-temperature. Atomic resolution STM and STS measurements were then carried out using a Unisoku STM system. Constant-current mode was used for all STM measurements. For STS, the lock-in technique was employed using a modulation voltage of 15 mV and frequency of 991 Hz.

Calculations were performed based on the DFT in conjunction with the projector-augmented wave potentials [22] as implemented in the Vienna Ab initio Simulation Package (VASP) [23,24]. The exchange-correlation potential was described through the Predew-Burke-Ernzerhof functional within the generalized gradient approximation formalism [25]. A plane wave basis set with a cutoff energy of 520 eV and K-mesh of 0.02 determined by a fine grid of Monkhorst-Pack method [26] in the Brillouin zone was found to produce good converged results. The atomic structures were relaxed using conjugate gradient algorithm until the forces on all unconstrained atoms were smaller than 0.005 eV/Å. A Vacuum layer of 20 Å along the $z$ direction was constructed to eliminate the interaction with spurious replica images [27]. The zero damping DFT-D3 method of Gimme [28] was used to account for long range vdW interaction between layers.

We note firstly that growth of crystalline Te on HOPG proceeded via the layer-by-layer mode as indicated by the streaky RHEED patterns as well as atomically flat surface morphology shown in STM micrographs. Figure 1a presents a topographic image of an as-grown Te sample for a nominal film thickness of ~ 20 Å. Large terraces delineated by atomic steps are clearly seen. Line profile measurements (see inset) reveal step heights of 4.0±0.2 Å. Atomic resolution image (Figure 1c) of such a surface reveals rectangular lattices instead of the hexagonal ones of the substrate and the in-plane lattice constants are measured to be $a = 4.26$ Å and $b = 5.42$ Å. They are consistent with the theoretically predicted *β*-tellurene that is schematically drawn in Figure 1b. It differs from the report of Ref [21], where a larger lattice of 5.93 × 4.42 Å$^2$ were noted for the Te layer grown on graphene. The latter is more in agreement with that of a bulk truncate of Te-I crystal along[$1\bar{0}10$]. We remark that a similarly larger lattice was also measured in our experiments from a thicker Te films (refer to Supplementary for a film $\geq 80$ Å thick grown on HOPG). Therefore, there could be a phase transformation from β-tellurene to Te bulk crystal as the film thickness increases. Further studies are however still needed to establish the thickness dependence of the phases as suggested by theory [11].

The STS measurements taken at flat terraces of the Te film are shown in Figure 2a for different thicknesses. Firstly, there is a clear density-of-state (DOS) gap, confirming that Te epifilms are semiconductors. The Fermi level is found below the mid-gap energy, suggesting that the sample is hole-doped. The origin of hole doping is likely from intrinsic defects, such as vacancies in the sample [29,30]. The experimental STS appear to agree well with the DFT calculated DOS shown in Figure 2b of the β-tellurene. The electronic gap of epitaxial Te is clearly film-thickness dependent, i.e., ~0.65 eV for the 5 tri-layer film (black line in Figure 2a), but ~0.58 eV and 0.53 eV for the 6 and 7 tri-layer films (red and blue lines in Figure 2a). An interesting feature in Figure 2a is that as the energy gap change happens predominantly at the VBM. The CBM stays almost constant at ~ 0.5 eV above the Fermi level. This reflects the very character of states at the VBM and CBM of ultrathin Te layers.

To gain further insight on the contributions at VBM and CBM, we carried out the DFT simulations for the partial charge densities of β-Te under different layer thicknesses. Figure 3(a) presents the derived partial density of states (PDOS) and Figure 3(b) plots the spatial distribution of charge densities at the energies of CBM and VBM for a single β-tellurene tri-layer (refer to Supplementary for the double and triple tri-layers cases and the band structures of β-Te for different thicknesses). It can be seen from both that the VBM states are mainly of the *p*-orbitals of the top and bottom layer Te atoms in the β-Te tri-layer, whereas the CBM states are mainly of the middle layer Te and thus confined within the β-Te tri-layer. As a result, stacking β-Te into multi-layer films will lead to inter-layer coupling of the VBM states but less influence on the CBM states. Consequently, one expects much more pronounced changes from the VBM than CBM as the film thickness varies, which agrees with the experiment. As a comparison, the partial charge density distributions of helical Te reveals greatly different characteristics (see Supplementary). As a result, the observation of the VBM change as a function of films thickness might be one manner of a β-Te film but a bulk-truncate in our experiment.

Finally, we have also acquired the STS across a step separating a 5-layer from a 6-layer terrace of epitaxial Te. The result is summarized in Figure 4a. As seen, apart from a discontinuous jump in energy bandgap at the step, corresponding to the thickness-dependent bandgap change as discussed above, no other obvious feature can be discerned. Particularly, there is no apparent band bending effect. Similar observation has been made at other steps separating different thickness layers. We remark that there could be a band-bending but over too small a spatial range to be detected by STM at the step edges. On the other hand, we note a similar observation made by Guo and co-workers where no band bending was seen across Te steps, despite an obvious bending observed at steps separating the substrate and monolayer Te islands [21].

One might expect the presence of edge states at steps due to broken translational symmetry and if so there could be a Fermi level pinning, resulting in the band-bending, for example at monolayer-to-bilayer steps of $MoSe_2$ [31,32]. However, our STS measurements at Te film steps reveal no mid-gap state (see also Figure 4b). So no edge states are present at least in the gap range of Te films. STM measurements (e.g., Figure 1c) reveal neither reconstruction of step-edge atoms, nor atom relaxation with the resolution limit of our measurements.

In conclusion, we have successfully grown Te films on HOPG by MBE and the in-plane lattice symmetry and lattice constants conform to the theoretically predicted β-tellurene for ultrathin layers. STS measurements reveal semiconducting properties of the Te epilayers and the energy bandgap decreases with increasing layer thickness. The gap variation occurs predominantly at the VBM while the CBM stays almost constant with changing film thickness. The latter is attributed to the different characters of the VBM versus CBM states as revealed by our DFT calculations for ultrathin β-Te layers. At steps, we observe no mid-gap edge state and band-bending effect.

# Acknowledgments

This work was financially supported in part by a Collaborative Research Fund (HKU9/CRF/13G) sponsored by the Research Grants Council (RGC), Hong Kong Special Administrative Region and by the National Natural Science Foundation of China (Grant nos. 61674053).

**Figures and Captions**

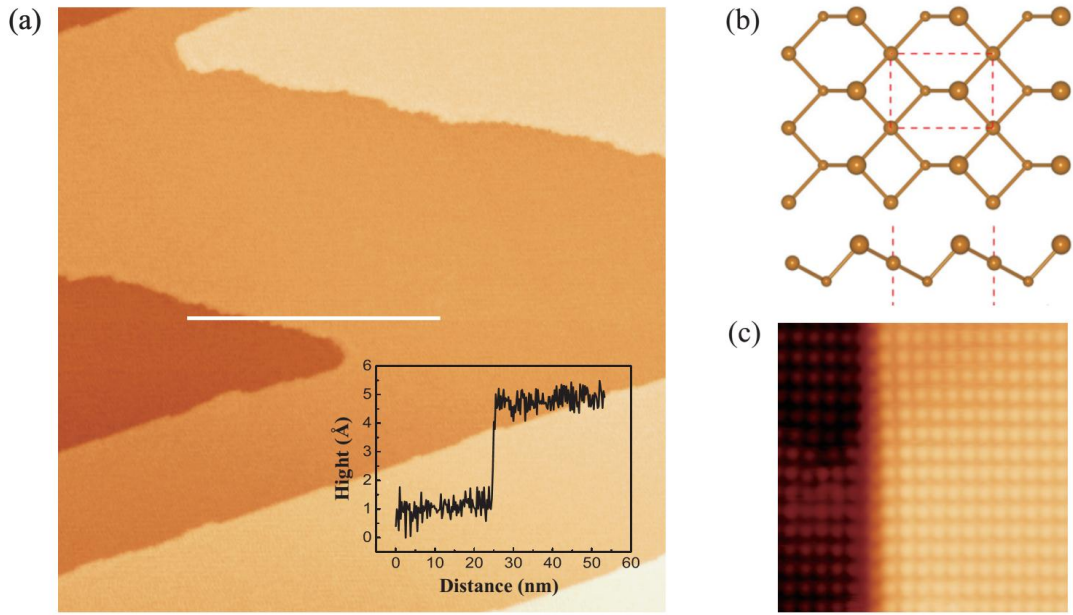

Figure 1. (a) Topographic image (size: 100×100 nm$^2$, sample bias: 1V) of an epitaxial Te film showing an atomically flat terraces separated by steps of height of ~ 4 Å. The inset presents a line profile taken along the white line drawn in the image. (b) A stick-and-ball model of β-tellurene layer viewed from top and side, respectively. (c) Atomic resolution STM image showing rectangular lattices (size: 8×8 nm$^2$, bias: 600 mV)

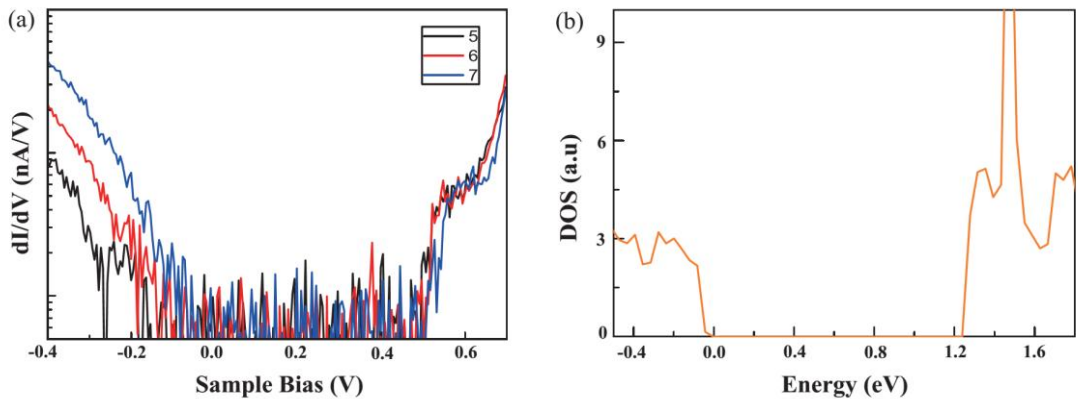

Figure 2. (a) STS of few-layer tellurene (black, red and blue are for 5, 6 and 7 layers, respectively), showing semiconductor property but thickness-dependence energy gaps. (b) DFT-calculated DOS of single-layer β-tellurene.

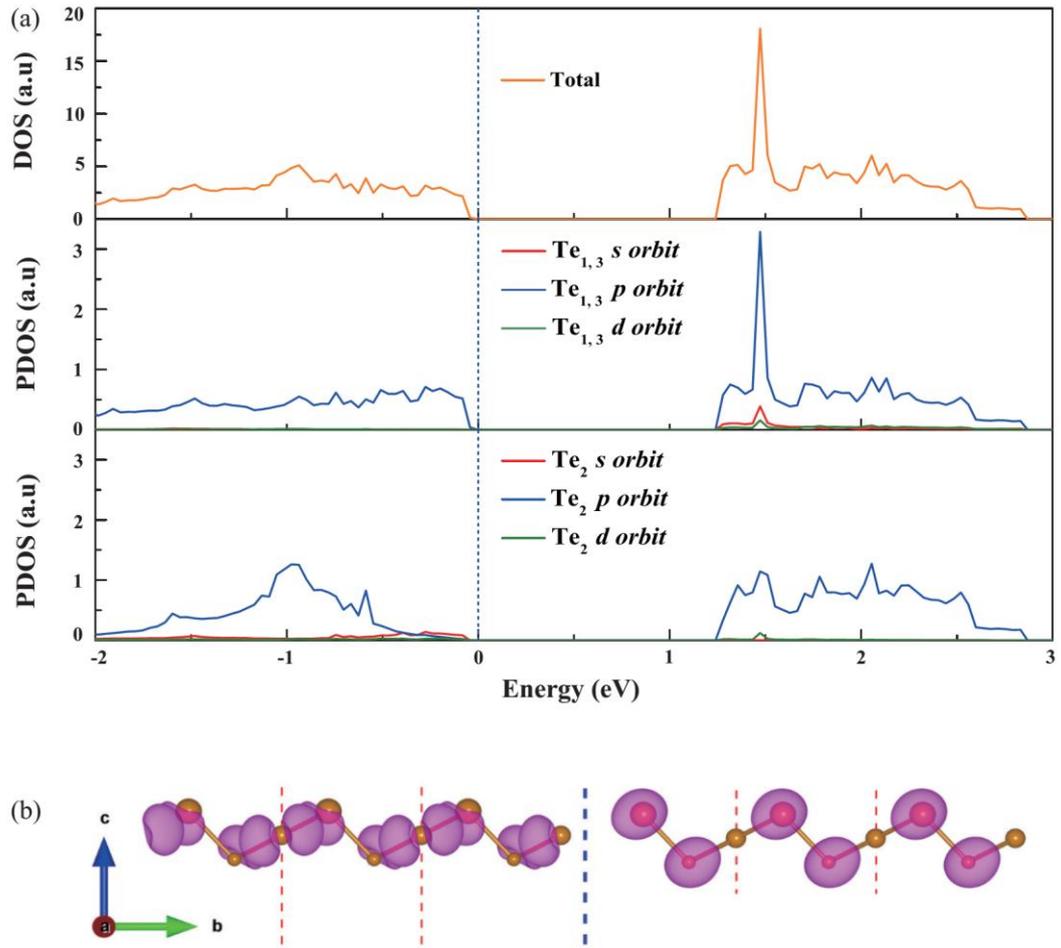

Figure 3. (a) DFT calculated PDOS, where Te$_{1,3}$ refer to atoms in the top and bottom layer of Te-tellurene, Te$_2$ refers that in the middle layer. (b) Spatial distribution for states at the CBM and VBM (isosurfaces are set at 0.04 $e$Å$^{-3}$), for a single-layer β-Te.

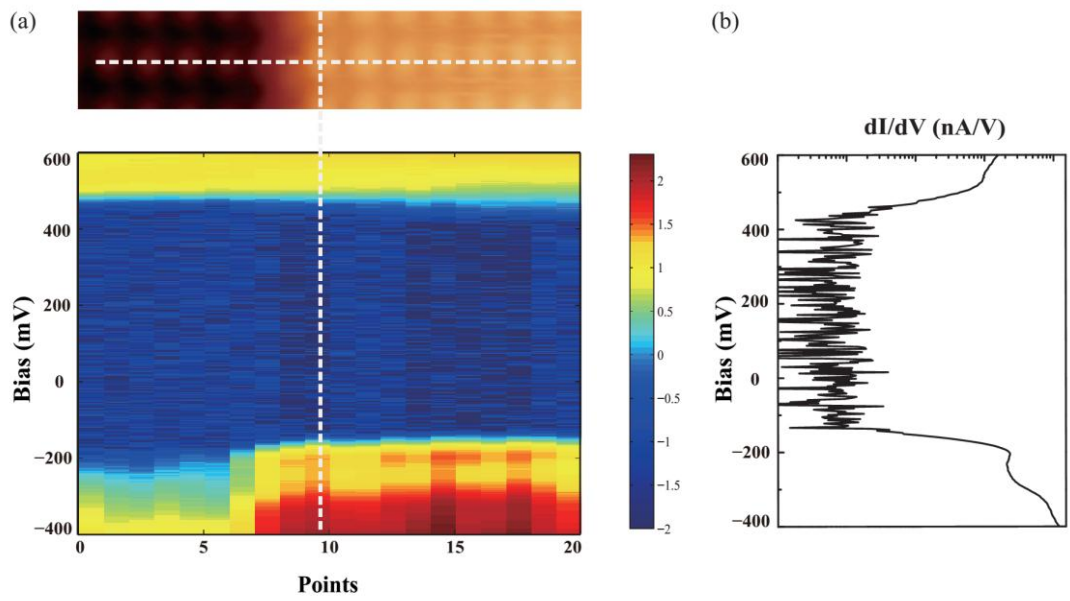

Figure 4. (a) Differential conductance spectra presented in the logarithm intensity map (bottom) taken across a 5-to-6 layer step (i.e., the horizontal dashed line in the STM image shown on top). (b) A spectrum taken over the step edge marked by the vertical dashed line.

**References**

[1] K. S. Novoselov, A. K. Geim, S. V. Morozov, D. Jiang, Y. Zhang, S. V. Dubonos, I. V. Grigorieva, and A. A. Firsov, Science **306**, 666 (2004).

[2] C. Berger *et al.*, Science **312**, 1191 (2006).

[3] A. C. Neto, F. Guinea, N. M. Peres, K. S. Novoselov, and A. K. Geim, Reviews of modern physics **81**, 109 (2009).

[4] P. Vogt, P. De Padova, C. Quaresima, J. Avila, E. Frantzeskakis, M. C. Asensio, A. Resta, B. Ealet, and G. Le Lay, Physical review letters **108**, 155501 (2012).

[5] L. Li, Y. Yu, G. J. Ye, Q. Ge, X. Ou, H. Wu, D. Feng, X. H. Chen, and Y. Zhang, Nature nanotechnology **9**, 372 (2014).

[6] A. J. Mannix *et al.*, Science **350**, 1513 (2015).

[7] B. Feng *et al.*, Nature chemistry (2016).

[8] F.-f. Zhu, W.-j. Chen, Y. Xu, C.-l. Gao, D.-d. Guan, C.-h. Liu, D. Qian, S.-C. Zhang, and J.-f. Jia, Nature materials **14**, 1020 (2015).

[9] B. Radisavljevic, A. Radenovic, J. Brivio, i. V. Giacometti, and A. Kis, Nature nanotechnology **6**, 147 (2011).

[10] X. Xu, W. Yao, D. Xiao, and T. F. Heinz, Nature Physics **10**, 343 (2014).

[11] Z. Zhu, X. Cai, C. Niu, S. Yi, Z. Guo, F. Liu, J.-H. Cho, Y. Jia, and Z. Zhang, arXiv preprint arXiv:1701.08875 (2017).

[12] F. A. Blum and B. C. Deaton, Physical Review **137**, A1410 (1965).

[13] Q. Wang, M. Safdar, K. Xu, M. Mirza, Z. Wang, and J. He, ACS nano **8**, 7497 (2014).

[14] Y. Xia, P. Yang, Y. Sun, Y. Wu, B. Mayers, B. Gates, Y. Yin, F. Kim, and H. Yan, Advanced materials **15**, 353 (2003).

[15] B. Mayers and Y. Xia, Advanced materials **14**, 279 (2002).

[16] M. Mo, J. Zeng, X. Liu, W. Yu, S. Zhang, and Y. Qian, Advanced materials **14**, 1658 (2002).

[17] Y. J. Zhu, W. W. Wang, R. J. Qi, and X. L. Hu, Angewandte Chemie **116**, 1434 (2004).

[18] C. J. Hawley, B. R. Beatty, G. Chen, and J. E. Spanier, Crystal Growth & Design **12**, 2789 (2012).

[19] M. Safdar, X. Zhan, M. Niu, M. Mirza, Q. Zhao, Z. Wang, J. Zhang, L. Sun, and J. He, Nanotechnology **24**, 185705 (2013).

[20] G. a. Tai, B. Zhou, and W. Guo, The Journal of Physical Chemistry C **112**, 11314 (2008).

[21] X. Huang, J. Guan, B. Liu, S. Xing, W. Wang, and J. Guo, arXiv preprint arXiv:1703.07062 (2017).

[22] G. Kresse and D. Joubert, Physical Review B **59**, 1758 (1999).

[23] G. Kresse and J. Furthmüller, Comp Mater Sci **6**, 15 (1996).

[24] P. E. Blöchl, Physical review B **50**, 17953 (1994).

[25] J. P. Perdew, K. Burke, and M. Ernzerhof, Physical review letters **77**, 3865 (1996).

[26] H. J. Monkhorst and J. D. Pack, Physical review B **13**, 5188 (1976).

[27] A. Bermudez, F. Jelezko, M. Plenio, and A. Retzker, Physical review letters **107**, 150503 (2011).

[28] S. Grimme, J. Antony, S. Ehrlich, and H. Krieg, The Journal of chemical physics **132**, 154104 (2010).

[29] A. Kolobov, Journal of non-crystalline solids **198**, 728 (1996).

[30] M. Kastner and H. Fritzsche, Philosophical Magazine Part B **37**, 199 (1978).

[31] C. Zhang, A. Johnson, C.-L. Hsu, L.-J. Li, and C.-K. Shih, Nano letters **14**, 2443 (2014).

[32] H. Liu, H. Zheng, F. Yang, L. Jiao, J. Chen, W. Ho, C. Gao, J. Jia, and M. Xie, ACS nano **9**, 6619 (2015).